\documentclass[vecphys]{svmult}

\usepackage{graphicx}        
\usepackage{multicol}        
\usepackage[bottom]{footmisc}

\usepackage{amsmath}

\usepackage{amsfonts}

\usepackage{cite}

\makeindex             

\begin{document}

\title*{What is special about water as a matrix of life?}
\titlerunning{What is special about water?}
\author{Lawrence R. Pratt\inst{1}, Andrew Pohorille\inst{2}, and D. Asthagiri \inst{1}}
\institute{Theoretical Division, Los Alamos National Laboratory, Los Alamos, NM  87545 USA
\texttt{lrp@lanl.gov} \& \texttt{dilipa@lanl.gov}
\and NASA, Ames Research Center,
Exobiology Branch, Moffett Field, CA 94035 USA \texttt{pohorill@raphael.arc.nasa.gov}}
%
%
\maketitle


\begin{abstract}
Water offers a large temperature domain of stable liquid, and the
characteristic hydrophobic effects are first a consequence of the
temperature \emph{insensitivity} of equation-of-state features of the
aqueous medium, compared to other liquids. On this basis, the known aqueous media and
conditions offer low risk compared to alternatives as a matrix to which 
familiar molecular biological structures and processes have adapted. The
current molecular-scale understanding of hydrophobic hydration is not
conformant in detail with a standard structural \emph{entropy
rationalization}. That classic pictorial explanation may serve as a
mnemonic, but isn't necessary. A more defensible view is that peculiar
hydrophobic effects can be comprehended by examination of engineering
parameters characterizing liquid water. LA-UR-05-3081
\end{abstract}

\setcounter{minitocdepth}{3}

\dominitoc

\section{Heuristic identification of solvent systems alternative to water as
a matrix of life}
\label{sec:intro}

The title of this discussion is a question that hints at a practical
interest in identifying alternative media for biomolecular processes and
structures.  In addition there is the curiosity to understand precisely
the special role that water plays in known biochemistry and biophysics.
Thus, we would like to understand the \emph{specialness} of liquid water
as a matrix of life \cite{AS72,Ball,franks2000} well enough to identify
close comparisons, and we would like to identify close comparisons in
order to refine our  understanding of the \emph{specialness} of water as
a matrix of life.

We are encouraged to embark on this discussion now because theories of
hydrophobic effects --- universally viewed as an essential feature of
aqueous biochemistry and biophysics --- have become more serious
over the past decade \cite{PrattLR:Molthe,HSA:2003}.  For many more
years than that there have standard views of hydrophobic effects
\cite{FRANKHS:FREVAE,AshbaughHS:Hydkac} that center on tetrahedrality as
a feature  of the local structure of liquid water, and upon entropic
contributions associated with constraints on the structure that might be
imposed by hydrophobic solutes.  That tetrahedrality concept, however,
has been ineffective in identifying alternatives.  It is defined on the
basis of known structures of liquid water, and then typically leads to a
view that there is no water but water.

Here we will refer to the classic views of hydrophobic effects based
upon that tetrahedrality concept as \emph{pictorial} theories because
those words were used initially \cite{pictoricaltheory}, and because
those views often don't proceed to a quantitative stage of molecular
science.  The recent progress on the problems of hydrophobic effects
\cite{Stillinger:73,Pohorille:JACS:90,Pratt:PNAS:92,Palma,%
HummerG:Anitm,Garde:PRL:96,Pratt:ECC,PrattLR:Molthe,PrattLR:Hydeam,%
HSA:2004,HSA:2003} lead in a different direction from the classic
pictorial theories.  The new direction avoids speculations on
structurally specific molecular mechanisms, and instead focuses on
engineering characteristics --- such as the equation of state of the
liquid water matrix --- in building a theoretical description.

From the perspective of that recent progress, the pictorial theories
that are side-stepped are  rationalizations typical of subtle but vital
chemical problems.  Though the alternative direction avoids simple
rationalization to achieve a fully defensible theory, it also leaves
open the sources of the necessary information. We use the context of
this discussion to collect some results on solvents that might offer
natural alternatives  to liquid water for astrobiology.  

A striking feature of the biophysical structures that we know is that
spatial organization is achieved by competing hydrophilic and
hydrophobic interactions \cite{TANFORDC:Thehea}.  It is natural then to
expect that an effective alternative matrix for life should support both
solvophobic \emph{and} solvophilic  interactions.   We comment on each
of these  issues in turn.

\subsection{Hydrophobic effects}

The most primitive conceptualization  of hydrophobic effects is just
that water and oil don't mix \cite{PrattLR:Hydeam,Mix:91}.  These phase
equilibria, together with molecular-scale hydrophobic-hydrophilic
amphiphilicity, result in spatially segregated mesoscropic structures
with obvious utilities \cite{TANFORDC:Thehea}.  The driving force for
oil-water phase separation is customarily viewed as a \emph{sticky}
interaction operating between hydrophobic species
\cite{KAUZMANNW:SOMFIT,CHANHS:THEPFP}, but not as specifically as
suggested by the archaic term `hydrophobic bond.'

\begin{figure}
\includegraphics[scale=0.5]{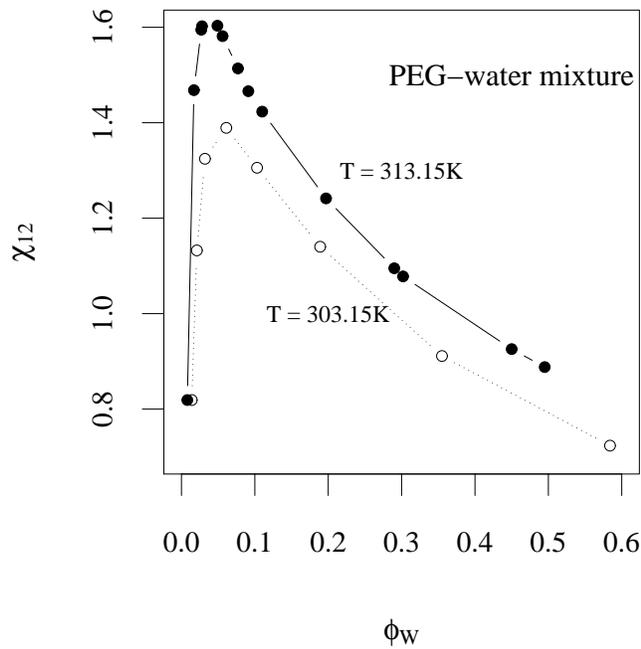}
\caption{Flory-Huggins interaction parameter for polyethylene glycol
water mixtures as a function of volume fraction of water for
temperatures 30~C (lower) and 40~C (upper).  This interaction parameter
is larger, favoring phase separation,  at the higher temperature,  an
expected behavior for hydrophobic effects.  The data are those of
\cite{BAEYC:REPOVL}; see \cite{BPP}.
}
\label{fig:chi}
\end{figure}

This phase equilibrium alone is not, however, the most challenging
aspect of the puzzle of hydrophobic effects.  Standard solution
thermodynamic studies establish that the suggested hydrophobic
stickiness becomes stronger as the temperature increases.  This point is
experimentally clear in the phenomenon of cold-denaturation of proteins
wherein cool, unfolded soluble proteins fold  {\em upon heating}
\cite{FRANKSF:LOWTUO,HATLEYRHM:DENOLD,FRANKSF:THETOP,Privalov:90}.
Another example of this behavior is shown in Fig.~\ref{fig:chi}.

It is inevitable that organized mesoscopic solution structures should
come apart at sufficiently high temperatures.  The fact that hydrophobic
--- stabilizing --- interactions should become stronger with increasing
temperature then stabilizes those structures over a broader range of
temperatures than if this peculiar temperature behavior of hydrophobic
interactions didn't obtain.  The temperature at which this increase of
hydrophobic attraction ceases is known as the entropy convergence
temperature, and is in  the neighborhood of 130~C at moderate
pressures; the upper-temperature limit of known life is in the
neighborhood of 120~C \cite{Kashefi}, so classic hydrophobic
interactions get stronger throughout that upper range of temperatures.

At the observed lower end of the temperature domain for metabolite
activity, perhaps near $-$20~C \cite{DemingJW:Psyapr,ThomasDN:OcesAs}, hydrophobic
stabilization of enzymes is expected to be more delicate
\cite{GeorletteD:Somlic}.  Though other conditions may not be
entirely the same in such a comparison, the systems of initial interest
are predominately aqueous systems in the temperature range is
$-$20~C $<T<$ 120~C.

The idea followed here is that this is a large temperature range.
Finding alternative solvents that might approach a temperature range
this large  is not expected to be easy.

A large temperature range is expected to be important because a living
organism is likely to require a multitude of physical and chemical
processes each with some temperature sensitivity.  Optimization of
individual such processes should be seen as secondary to satisfactory
compromises of performance among many processes.  An  alternative medium
that offers a wide range of possibilities for such compromises should be
a preferred choice as an alternative for the known matrix of life.

Viewed contrapositively, a medium in which a significant fraction of
required processes could be seriously degraded by  a thermal
excursion would not be expected to be a satisfactory matrix of  life. 
Then the temperature maxima and minima that abound in physical
properties of aqueous solutions --- the maximum of the density in the
liquid is one example, and the minimum in the compressibility shown in
Fig.~\ref{fig:c22} is another example --- the significance of these
maxima and minima is first that the properties of liquid water have a
small net variation in an extended temperature range.

\subsection{Hydrophilic effects}

\begin{figure}
\begin{center}
\leavevmode
\includegraphics[scale=1.0]{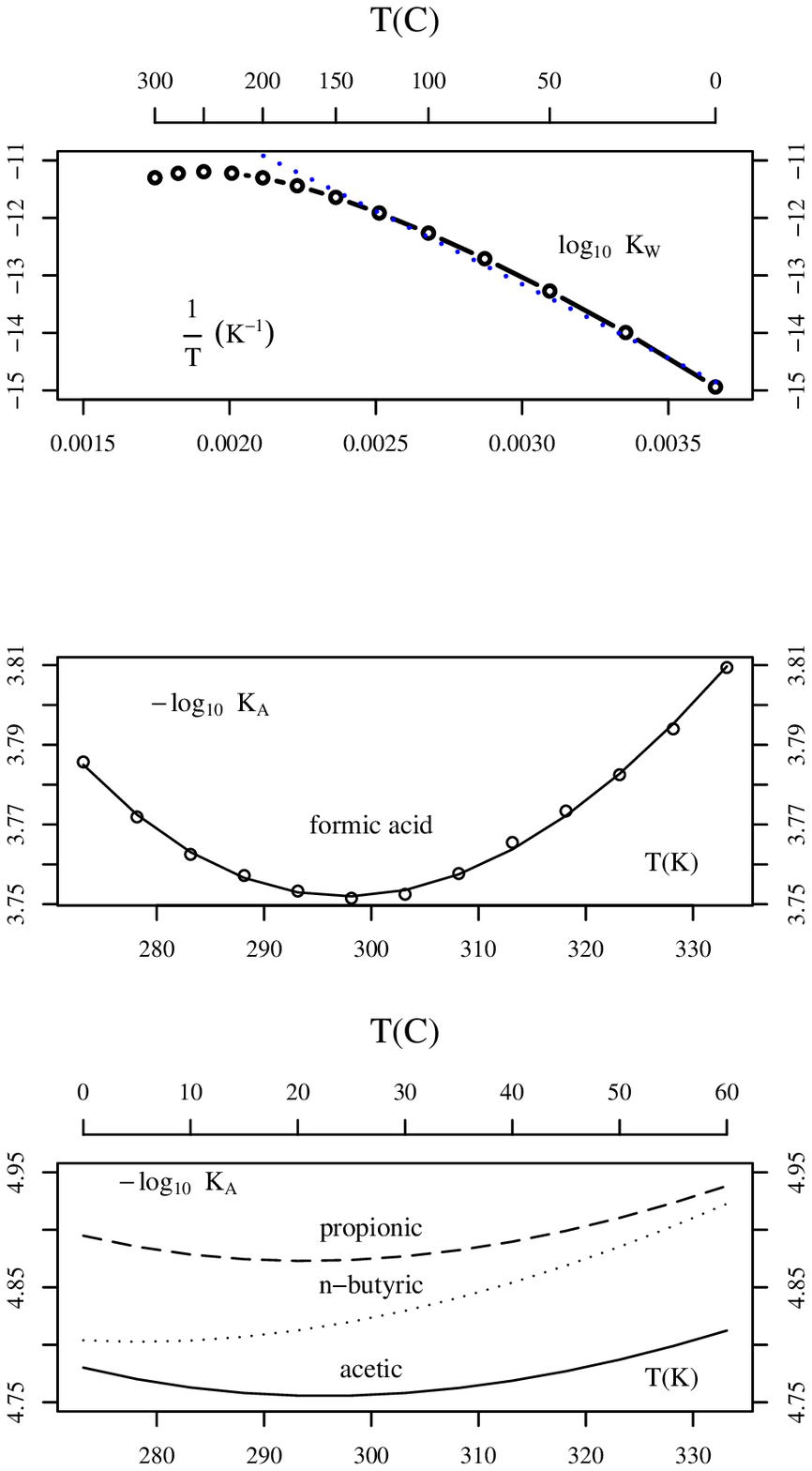}
\end{center}
\caption{\emph{Upper:} the ion product $K_\mathrm{W}$ = $\left\lbrack
\mathrm{H}^+\right\rbrack \times \left\lbrack
\mathrm{OH}^-\right\rbrack$ with concentrations in units of mol/l, for
liquid water along the vapor saturation curve \cite{Marshall:81}.  The
dotted curve is a  model of the form $- \log_{10} \left\lbrack
\mathrm{H_2O} \right\rbrack + a + b/T$ fitted to the data below 150~C.
The analogous ion product for liquid  ammonia, $\left\lbrack
\mathrm{H}^+\right\rbrack \times \left\lbrack
\mathrm{NH}_2^-\right\rbrack$, is reported to be about 10$^{-33}$
\cite{Life}.  \emph{Lower:} From \cite{HS43}.
Except for the $n$-butyric case, the magnitudes of the standard
enthalpies for these dissociation reactions are substantially less than
the thermal energy here.}
\label{fig:pKW}
\end{figure}

If traditional
hydrocarbon-dominated  organic molecules are the carriers of one of
those categories of interaction, that seems to leave classic
electrostatic and chemical interactions for the other category
\cite{Pratt:ECC}.  Indeed, water is a chemically active liquid.
Water is involved  in standard
acid-base chemistry, both  through direct molecular participation, and
indirectly by auto-ionization:
\begin{eqnarray}
\mathrm{H_2O} \rightleftharpoons \mathrm{HO}^- + \mathrm{H}^+  
\end{eqnarray}
This underlies the buffering that is essential to familiar biochemistry,
and moderates changes in charge states of soluble macromolecules. 
Fig.~\ref{fig:pKW} shows temperature variation of the ion product
$K_\mathrm{W}$ = $\left\lbrack \mathrm{H}^+\right\rbrack \times
\left\lbrack \mathrm{OH}^-\right\rbrack$. It increases significantly
with increasing temperature along the saturation curve as both the
density and the dielectric constant decrease.

In the context of protein folding studies some years ago, Klotz
\cite{KlotzIM:Parcwt} drew attention to the fact that $pK_\mathrm{A}$ of
simple carboxylic acids show extrema curiously similar to the others we
have been discussing;  this is exhibited in Fig.~\ref{fig:pKW}.

Familiar aqueous biochemistry exploits standard acid-base chemistry and
mobile ions in solution.  For example, the interiors of cells have a
negative electrical potential relative to the exterior. This is because
electrical potential differences couple to proton shuttling and ATP
synthesis. Regulated electrical communication seems necessary. Mobile
ionic solutes serve that purpose.

\begin{figure} \begin{center} \leavevmode
\includegraphics[scale=0.9]{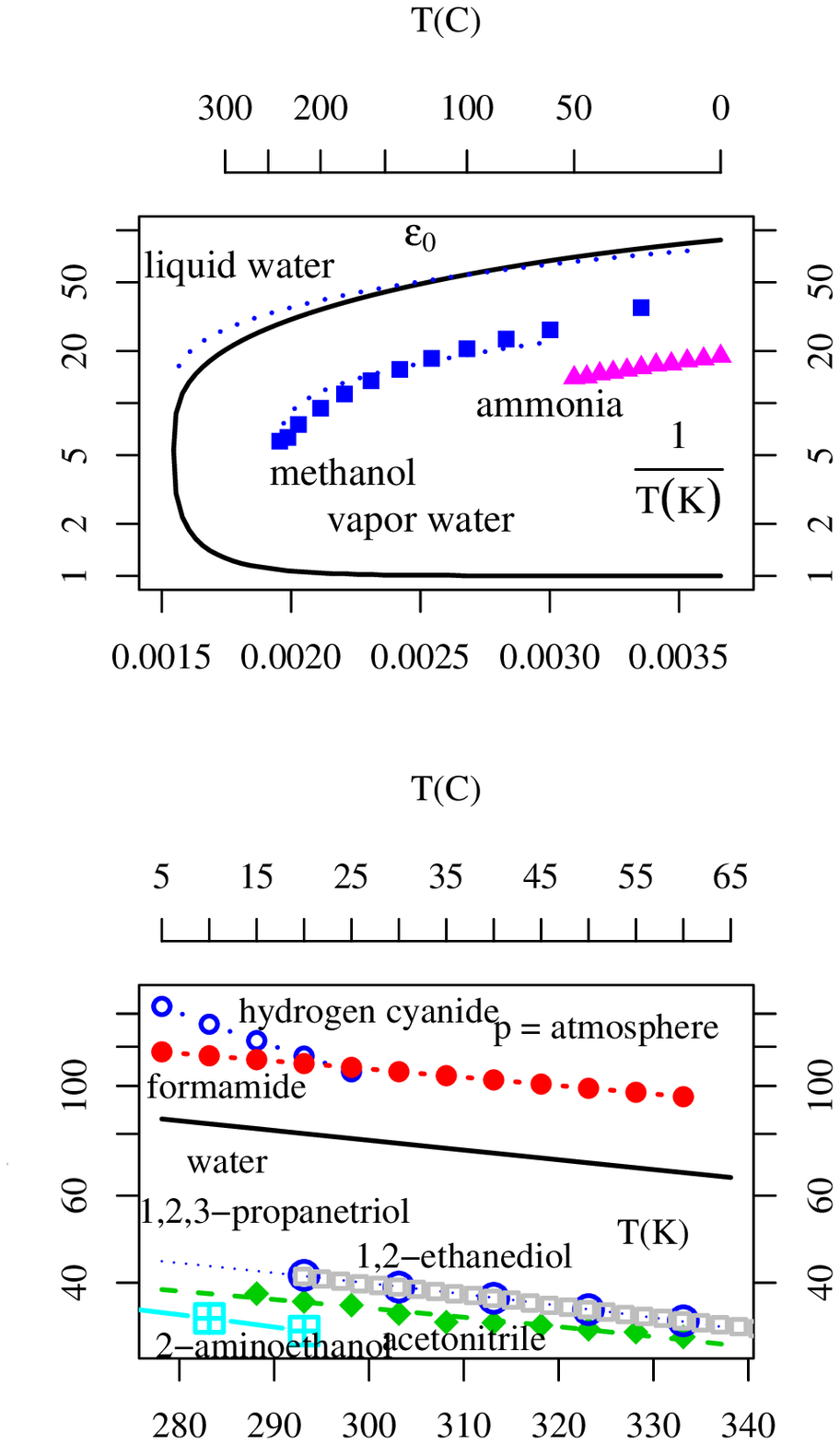} \end{center} 
\caption{\emph{Upper:} the static dielectric constant of fluid water
along the liquid-vapor coexistence curve  \cite{Uematsu:80}, and liquid
methanol \cite{Takaaki:04} and liquid ammonia \cite{Billaud:75}. The
blue dotted curves are fitted models $\propto\rho_\mathrm{liq}/T$,
suggesting that the qualitative behavior is simply understood on the
basis of equation of state variables. Measurement of the dielectric
constant of \emph{liquid} H$_2$S as a function of temperature has been
sketchy; but the value at its normal freezing point, about 9.3, gives a
typical magnitude.  The static dielectric constant of liquid ammonia is
about 23 for the lowest temperature liquid state. \emph{Lower:} Static
dielectric constants as a function of $T$ at ambient pressures for
several high-dielectric-constant liquids; see \cite{LBb}. Hydrogen
cyanide might be the highest-dielectric-constant molecular liquid, but
with the most rapid decrease with $T$ through these thermodynamic states
\cite{coates}.  The electrical conductivity of HCN is lower than that of
liquid water.  Formamide has a higher dielectric constant than liquid
water here, and that dielectric constant doesn't decrease faster than
does the dielectric constant of water \cite{dunn}.  A variety of amides,
and also urea and thiourea, are likely to exhibit similarly high
dielectric constants.  Liquids with static dielectric constants above 30
are fairly common, and some of those cases are shown.} \label{fig:e1bar}
\end{figure}

Ionic dissolution and dissociation in liquid water is associated with
its high dielectric constant.  Fig.~\ref{fig:e1bar} shows that this
dielectric constant is above 50 for temperatures  below 120~C;  these
are high values, and so high that the variation of $\varepsilon_0$ in
the liquid phase here is not a dominating concern.  Relative variations
of hydration free energies will depend on
$\delta\varepsilon_0/\varepsilon_0$. The observed high-values of the
dielectric constant means that salts are satisfactorily soluble in
liquid water, and that good solubility will not be critically sensitive
to variations of solvent quality.  Dissolution of minerals extends the
$\left(T,p\right)$ domain of the liquid phase.

Liquids with dielectric constants as high as those of liquid water are
not usual; results for a selection of such cases are shown in
Fig.~\ref{fig:e1bar}.

\section{Current theory of hydrophobic effects}

Why should hydrophobic attractions become stronger at higher
temperatures? This has always been a contentious issue for molecular
theories \cite{BENNAIMA:ONAOS,ROSSKYPJ:BENIIA}. But a rationalization
based upon a structural mechanism has always been standard
\cite{FRANKHS:FREVAE}: water molecules contacting  hydrophobic groups
are considered to be ordered, suffering an entropy penalty
\cite{SilversteinKAT:Hydsmw}.  When hydrophobic groups are buried by
phase  segregation or folding, contacting water  molecules are
\emph{released to the bulk,} regaining the penalized entropy.  The
demixed or folded case has a higher entropy for that reason, and is
the stable outcome at higher temperatures.  Viewed  in the opposite
temperature direction, this suggests that  cold-denaturation works by
water molecules prying-open folded, soluble protein molecules
\cite{Paulaitis:02}.

This \emph{entropy rationalization} might have a fractional truth.  But
it has never been developed in a  conclusive way and proved. An 
appreciation of this gap between molecular theories and common
rationalizations seems to be growing more widely \cite{SoperAK:Exceas}.

Furthermore,  experimental counter-examples are readily available,
simple, and troubling. A particularly clear counter-example is due to
Friedman \& Krishnan \cite{Friedman-Krishnan}: the sum of the standard
hydration entropies of K$^+$(aq) and Cl$^-$(aq) is about \emph{twice}
the standard hydration entropy of Ar(aq).  The case of methanol as
solvent is qualitatively different.  If hydrophobic effects are
conceptualized on the basis of hydration entropies and specific
hydration structures, this is paradoxical: according to the measured
entropies Ar(aq) + Ar(aq) is  about as hydrophobic as K$^+$(aq) +
Cl$^-$(aq), but the hydration structures neighboring K$^+$(aq),
Cl$^-$(aq), and Ar(aq) are clearly  qualitatively different, both in
structures observed and  numbers  of water molecules affected. A
speculation how this might happen was offered  recently \cite{CPMS},
but couldn't be considered yet proved.

This example should not be interpreted to suggest that K$^+$(aq) +
Cl$^-$(aq) is hydrophobic.  The point is that the classic \emph{entropy
rationalization} is not a necessary explanation why Ar(aq) is
hydrophobic.  

The proved theories of classic hydrophobic effects
\cite{Stillinger:73,PrattLR:Molthe,AshbaughHS:Hydkac,HSA:2003} --- 
leaving aside  the cases of ionic solutes or of hydrophobic
effects in the context of amphiphilic solutes --- establish something
more subtle and more general than the structural \emph{entropy
rationalization} discussed above: the entropic effects that are
characteristic of hydrophobic phenomena are closely tied to
peculiarities of the equation of state of  liquid water. This is the
engineering perspective that we noted at the outset of this discussion.
\begin{figure}
\includegraphics[scale=0.58]{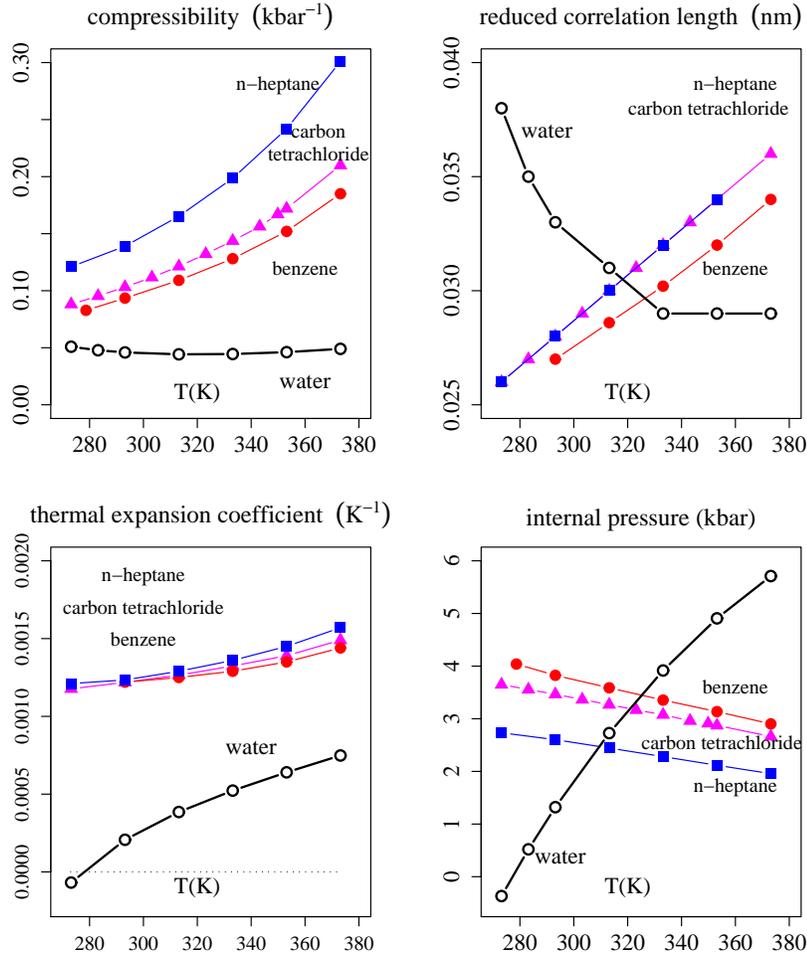}
\caption{
Key thermodynamic parameters associated with the modern theory of
hydrophobic effects, variation with temperature along the vapor
saturation curve. \emph{Upper-left:} compressibility, $\kappa_T\equiv\left(\partial\ln
\rho/\partial p\right)_T$.  Water is stiffer than comparable organic
solvents, and doesn't appreciably soften in this temperature range.
\emph{Lower-left:} thermal expansion coefficient,
$\alpha_\sigma\equiv-\left(\partial\ln\rho/\partial T\right)_\sigma$. Water expands less
rapidly with temperature than comparable organic solvents.
\emph{Upper-right:}  the product of the liquid-vapor surface tension 
and the liquid compressibility, $\gamma_{\mathrm{lv}} \times
\left(\partial\ln \rho/\partial p\right)_T$: the reduced correlation
length of \cite{EGELSTAF*PA:Liqstn}. The temperature
dependence of this correlation length for water is qualitatively
different from these organic solvents. \emph{Lower-right:}  The internal
pressure,  $\left(\partial U/\partial V\right)_T$. The hydrocarbon liquids
qualitatively conform to the van~der~Waals expectation that this should be
proportional to $\rho^2$, but water is qualitatively different.
Data from \cite{Rowlinson} and \protect{\cite{Jasper,LBa}}.
}
\label{fig:c22}
\end{figure}

\subsection{Equation-of-state features that govern primitive hydrophobic effects}
Fig.~\ref{fig:c22} then grounds our description of these engineering
aspects of the present problem \cite{PrattLR:Hydeam,AshbaughHS:Hydkac}.
The \emph{Upper-left} panel there shows measured compressibilities for liquid
water and several organic solvents.  Water is qualitatively stiffer. 
Current theory \cite{PrattLR:Molthe,HSA:2003} suggests that the low
solubility of inert gases in liquid water is first a reflection of this
stiffness.  Notice that this stiffness is only weakly sensitive to
temperature along the vapor saturation curve, in contrast to the
stronger temperature dependence of the compressibilities of the other
solvents.

Compressibilities of 150 liquids at ambient conditions of $T$ = 298.15~K
and $p$ = 1~bar were surveyed in \cite{marcus:97}. In that collection,
liquid water does indeed have a low compressibility;  glycerol had the
lowest compressibility in that listing.  Working from  lowest
compressibilities upward  those liquids are ordered as glycerol,
diethylene glycol, ethylene glycol, formamide, 2-aminoethanol, 1-4
butanediol, 1-5 pentanediol, and water, with the last two noted cases
tied. This ordering suggests that all of the named liquids would be of
interest for further study in an astrobiology context. Formamide not
only is less compressible than liquid water, it also has a
compressibility \emph{minimum} near 25~C at low pressure \cite{easteal}. 

The \emph{Lower-left} panel of Fig.~\ref{fig:c22} makes a similar comparison
of the thermal expansions of our standard organic solvents along the saturation curve.
We are principally interested in the circumstances that the density
decreases with increasing temperature, but that decrease is
qualitatively slower for liquid water than for the other solvents shown.
Current theory \cite{PrattLR:Hydeam,AshbaughHS:Hydkac} suggests that the
inverse of the thermal expansion coefficient $-\left(\partial
T/\partial\ln\rho\right)_\sigma$ should be considered a characteristic
temperature for hydrophobic hydration;  in the most fortunate cases
\cite{PrattLR:Hydeam,AshbaughHS:Hydkac}, it is --- roughly but clearly
--- a small multiple of the \emph{entropy convergence temperature}
observed experimentally with the solubilities of inert gases in water
\cite{Baldwin:PNAS:86,Baldwin:PNAS:92}.  Where this entropy
convergence behavior is well-developed, the entropy convergence
temperature is the temperature at which the hydrophobic stickiness stops
increasing in strength with increasing temperature. Following the
interpretation of $-\left(\partial T/\partial\ln\rho\right)_\sigma$ as
an important temperature scale, that temperature --- see
Fig.~\ref{fig:c22} --- is higher for liquid water than for the
other solvents shown.

It is often guessed that the surface tensions  associated with water
interfaces should be involved in the description of  hydrophobic effects
\cite{AshbaughHS:Hydkac,SHARPKA:RECTMO}. A nice way to consider what
might be the implications of liquid-vapor interfacial tensions,
$\gamma_{\mathrm{lv}}$, is to consider the product $\gamma_{\mathrm{lv}}
\times \left(\partial\ln \rho/\partial p\right)_T$; this is a
\emph{reduced correlation length}
\cite{AshbaughHS:Hydkac,EGELSTAF*PA:Liqstn}, and is shown for the same
cases in the \emph{Upper-right} panel of Fig.~\ref{fig:c22}. For the 
organic solvents the tensions decrease and the  compressibilities 
increase with increasing temperature.  In  those cases, the
compressibility-increases dominate, and the correlation lengths
increase.  For liquid water,  in contrast, the compressibility scarcely
changes, and the interfacial tension decreases with increasing
temperature.  Thus, this correlation length decreases with increasing
temperature.  Recent theories suggest that a multiple --- roughly  a
factor of ten --- of this reduced correlation length identifies a
length-scale on which macroscopic considerations can be directly
exploited in proved theories \cite{AshbaughHS:Hydkac}.

There are two points of interpretation of the results for the reduced
correlation length. The first point is that though liquid water has a
relatively high interfacial tension with its vapor,  when viewed in this
scale the magnitudes are not unusual.   It is the temperature dependence
that is unusual, and that unusual temperature dependence is mostly
associated with the compressibility.  The second point of interpretation
is that the contraction of this correlation length with increasing
temperature naturally would  be viewed as a break-down of the more open
architecture of liquid water at lower temperatures.  This correlation
length is apparently a sensitive diagnostic of this break-down behavior;
the more detailed current theory also predicts a contraction with
increasing temperature of a corresponding natural length scale
\cite{AshbaughHS:Hydkac}.

The final frame of Fig.~\ref{fig:c22}, \emph{Lower-right}, shows the
internal pressure,  $\left(\partial U/\partial V\right)_T$.  Following a
van~der~Waals model $\left(\partial U/\partial V\right)_T\approx
a\rho^2$, with $a$ the van~der~Waals parameter describing the effects
of attractive intermolecular interactions on the equation of state. This
contribution stabilizes the liquid at low pressure. Water is clearly
qualitatively different from the organic solvents in this respect.
Because of the expectation of a van~der~Waals model, the behavior seen
for liquid water might suggest that this liquid is becoming better bound
at higher temperatures, definitely counter-intuitive and a paradoxical
view.

This example illustrates how a counter-intuitive temperature dependence
can arise.  This internal pressure can be expressed as 
\begin{eqnarray}
 \left(\frac{\partial U}{\partial V}\right)_T =  T \left(\frac{\alpha_p}{\kappa_T}\right) - p,
\label{eq:pint}
\end{eqnarray}
where $\alpha_p\equiv-\left(\partial\ln\rho/\partial T\right)_p$ here is
quantitatively only slightly different from the $\alpha_\sigma$ of
Fig.~\ref{fig:c22}, and the compressibility $\kappa_T$ was defined there
also.  For the case of liquid water, $\kappa_T$ is small, and  since
$\alpha_p$ is also comparatively small the density changes are small.
The change of $\kappa_T$ with $T$ is thus also small, the ratio
${\alpha_p}/{\kappa_T}$ is, above 4~C, then a substantial factor with
weak $T$ dependence, and the $p$ contribution  of Eq.~\eqref{eq:pint} is a
secondary magnitude.  The formal multiplier of $T$ in
Eq.~\eqref{eq:pint} then causes a steep increase with increasing
temperature.  Thus the literal description of the counter-intuitive
behavior is that with increasing temperature
\begin{enumerate}
\item the density is decreasing but only
slowly,
\item  the low
compressibility is scarcely changing, 
\item and the kinetic energy --- the  explicit factor of $T$ --- is
increasing.  
\end{enumerate}
This combination produces the counter-intuitive result which is still an
amazing trick, but at least somewhat clearer.  The counter-intuitive
temperature dependences of hydrophobic effects such as that shown in
Fig.~\ref{fig:chi} eventually acquire a practically identical
explanation \cite{Garde:PRL:96,PrattLR:Molthe}. Consequently, simple
theories of hydrophobic effects that incorporate equation-of-state
information properly --- and scarcely anything else --- can be
qualitatively correct in describing many of these unusual temperature
dependences.  

Note that this distinctive temperature dependence continues to
temperatures near the currently known upper-temperature limit of life.

\begin{figure}
\begin{center}
\leavevmode
\includegraphics[scale=0.8]{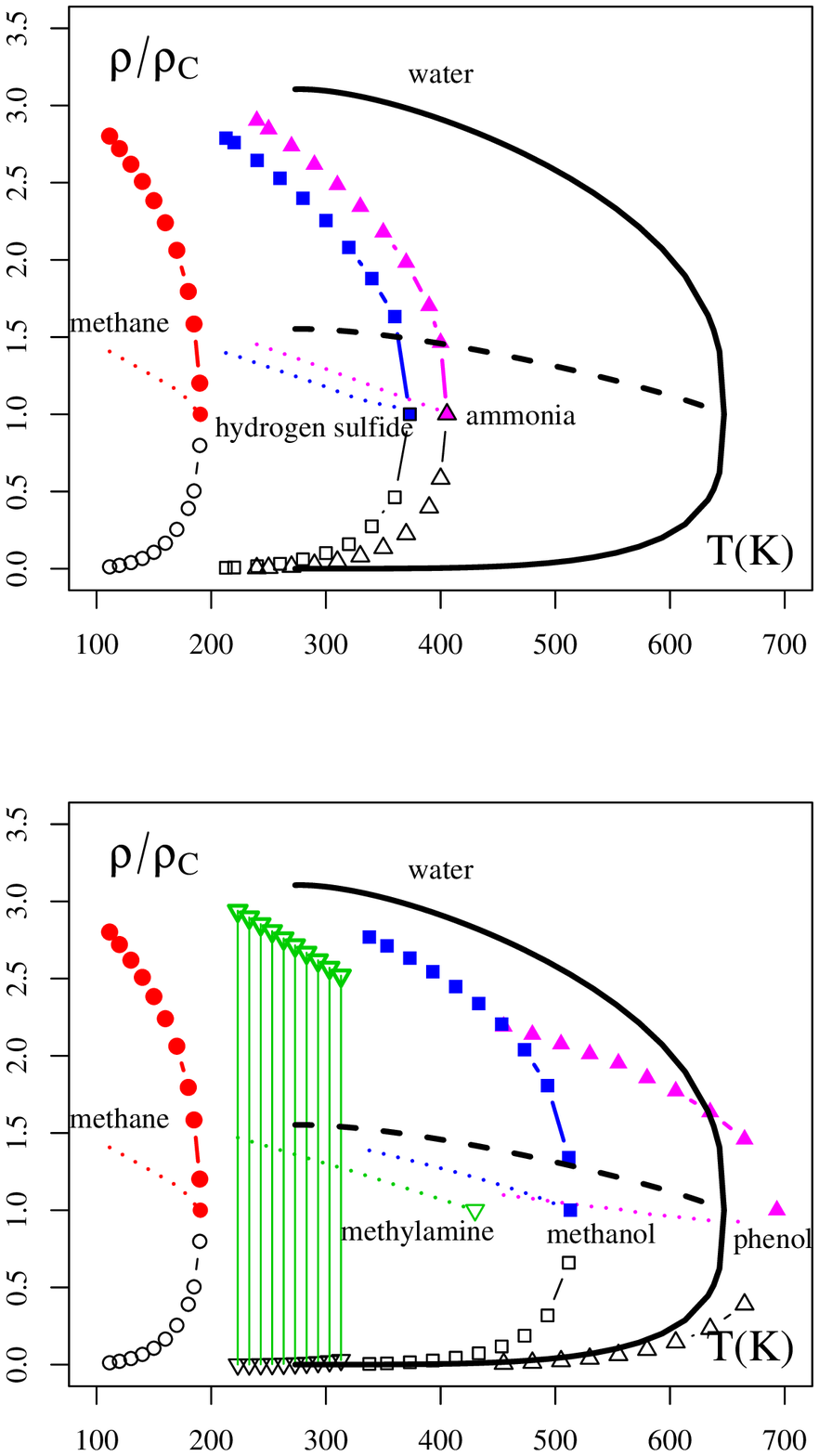}
\end{center}
\caption{Liquid-vapor coexistence densities for several fluids compared to
liquid water;  from \cite{Beaton:89} except for the methylamine case
which comes from \cite{Matheson80}.  The vertical axis is the  density
relative to the critical density in each case.  The temperature domain
of the liquid water phase is larger than the other cases here. The
variation of the density of liquid water is slower than the other
cases, consistent with  Fig.~\ref{fig:c22}.}
\label{fig:coex}
\end{figure}

\subsection{Water compared to other possibilities}

We can take a broader view of these characteristics of  liquid water by
comparison with some other fluids that might be considered in the
context of  astrobiology.  Fig.~\ref{fig:coex} shows phase diagrams for
a selection of interesting cases.  The oft-noted point that water has a
high critical temperature compared to other small molecule liquids is
emphasized by these results.  That critical temperature is well above the
temperatures of first interest here, so it is probably not specifically
relevant.  But water achieves this high critical temperature by
exhibiting a relatively long temperature domain for the liquid phase;
and then the density of liquid water decreases comparatively slowly with
increasing  temperature.  This is consistent with the results of
Fig.~\ref{fig:c22}.

\section{Discussion}

In view of the intricate beauty of molecular bioscience, a common
attitude is the biomolecular structures and processes are delicately
tuned to the aqueous medium and conditions we know.  The discussion here
suggests that the familiar molecular biology has adapted to the known medium
and conditions that offer  low risk compared to alternatives.  Water
offers a large temperature domain of stable liquid, and the
characteristic hydrophobic effects are first correlated with the
temperature \emph{insensitivity} of engineering properties of
the aqueous medium, compared to other liquids.

This suggests that molecular biological structures and processes often could
be isolated to function satisfactorily in alternative milieu.  But it is
likely that the conditions would require more careful control than  in
the original aqueous setting.  Transplanting successively more
biochemical processes to function satisfactorily at the same time in the
same alternative milieu and conditions naturally should be  considered 
as increasingly hazardous.

\section{Conclusions}

The current molecular-scale understanding of hydrophobic hydration is
not conformant in detail with a standard structural \emph{entropy
rationalization}.  That standard pictorial explanation may serve as a
mnemonic, but isn't necessary. A more defensible view is that peculiar
hydrophobic effects can be comprehended by examination of engineering
parameters characterizing liquid water. This is even simpler than the
standard pictorial theory.

\bibliographystyle{SV}
\bibliography{Outline,book.01}

%



\end{document}